\documentclass[11pt]{article}

\usepackage[]{graphicx,float,latexsym,times}
\usepackage{amsfonts,amstext,amsmath,amssymb,amsthm}
\usepackage{amsmath,mathrsfs}
\usepackage{caption2}

\long\def\symbolfootnote[#1]#2{\begingroup%
\def\thefootnote{\fnsymbol{footnote}}\footnote[#1]{#2}\endgroup}

\usepackage{titlesec} 

\titleformat{\section}{\large\bfseries}{\thesection.}{.5em}{}
\titlespacing*{\section}{0pt}{*3}{*2}
\titleformat{\subsection}{\normalfont\bfseries}{\thesubsection.}{.5em}{}
\titlespacing*{\subsection} {0pt}{*3}{*2}
\titleformat{\subsubsection}{\normalfont\bfseries}{\thesubsubsection.}{.5em}{}
\titlespacing*{\subsubsection} {0pt}{*3}{*2}

\theoremstyle{plain} 
\newtheorem{theorem}{Theorem}[section]

\theoremstyle{definition} 

\newtheorem{remark}{Remark}[section]
\newtheorem{proposition}[theorem]{Proposition}

\numberwithin{equation}{section} 

\usepackage[authoryear]{natbib}
\begin{document}

\title{\textbf{\Large Group sequential hypothesis tests with variable group sizes: optimal design and performance evaluation}}

\date{}

\maketitle


\author{
\begin{center}
\vskip -1cm

\textbf{\large Andrey Novikov}

 Metropolitan Autonomous University, 
Mexico City, Mexico

\end{center}
}

\symbolfootnote[0]{\normalsize Address correspondence to A. Novikov,
 Universidad Aut\'onoma Metropolitana, Unidad Iztapalapa, Avenida Ferrocarril  San Rafael Atlixco, 186, col. Leyes de Reforma 1A Secci\'on, C.P. 09310, Cd. de M\'exico, Mexico,
 E-mail: an@xanum.uam.mx}

{\small \noindent\textbf{Abstract:} 
In this paper, we propose a computer-oriented method of construction of optimal group sequential hypothesis tests with variable group sizes. 
In particular, for independent and identically distributed observations we obtain the form of optimal group sequential tests which turn to be a particular case of sequentially planned probability ratio tests \cite[SPPRTs, see][]{Schmitz}.
Formulas are given for computing the numerical characteristics of general SPPRTs, like error probabilities, average sampling cost, etc.
A numerical method of designing the optimal tests  and   evaluation of the performance characteristics is proposed, and computer algorithms of its implementation are developed.
For a particular case of sampling from a Bernoulli population, the proposed method is implemented in R programming language,  the code is available in a public GitHub repository.
The proposed method is compared numerically with other known sampling plans.
}
\\ \\
{\small \noindent\textbf{Keywords:} sequential analysis;
sequentially planned procedure;
hypothesis test;
optimal sampling; optimal stopping
 }
\\ \\
{\small \noindent\textbf{Subject Classifications:}  62L10, 62L15, 62F03, 60G40
}

\section{INTRODUCTION} \label{Intro}

Group sequential hypothesis tests with variable group sizes have been proposed as  a theoretical framework for some practical situations when sequential statistical methods are applied to sampling  data in groups.
 In many occasions, the overall cost of sampling in groups includes, additionally to  the unitary cost   of any collected data item, also some set up cost  related to the group, in which case the classical one-per-group sequential plans \citep[see][]{waldwolfowitz} may not be optimal with respect to the total cost \citep{Cressie}. \cite{Ehrenfeld} used the general principle of dynamic programming for obtaining
 the form of optimal group sequential sampling plan in the  Bayesian set-up, with arbitrary cost function. \cite{Schmitz} proposed a unified approach to  sequentially planned statistical procedures based on optimal stopping with respect to a partially ordered ``time line''. For the sequentially planned hypothesis tests  he proposed a general structure of testing  procedures, called sequentially planned probability ratio tests (SPPRTs), where the rule responsible for sampling may vary depending on particular problem settings.

 For the problem of testing of two simple hypotheses, \cite{SchmegnerBaron} studied a class of SPPRTs under the assumption that the log-likelihood ratio takes its values on a lattice. In this case, they obtained analytical expressions for characteristics of the SPPRTs and, in the particular case of symmetric hypotheses about the success probability in the Bernoulli model, evaluated them for various  sampling plans known from the literature, aiming at minimisation of the  average cost of observations.

Another large area of applications related to the group sequential testing is related to clinical trials. The variable-group-size plans are called adaptive in this context \citep{Dragalin}. In \cite{Eales}, the dynamic programming principle is used for obtaining optimal adaptive group sequential testing procedures in the normal model.

In this paper, we propose a computer-oriented method of construction of optimal group sequential  tests with variable group sizes. The basic idea of the proposed method is to use a grid interpolation scheme in the backward induction equations. We develop a complete set of computer algorithms for their design and performance evaluation. For the case of Bernoulli observations, we implement the algorithms in R programming language \citep{R} and numerically  compare the  obtained plans with those of \cite{SchmegnerBaron}. The relative efficiency of the optimal test with respect to the  one-stage plan with the same levels of error probabilities is evaluated. 

Another application we consider is with respect to the optimal adaptive group sequential tests for Phase II clinical trials based on binary outcomes. In this example we numerically compare the performance of our optimal plans with those proposed by  \cite{Fleming1982}.

In Section 2, general results on optimal group sequential tests with variable group sizes are summarized.
In Section 3,  optimal group sequential tests with variable group sizes for independent and identically distributed observations are characterized (which turn out to be of SPPRT type) and formulas for performance characteristics are obtained.
In Section 4, a numerical method of optimal sequential planning and evaluation of  performance characteristics  is proposed. Numerical examples are presened.
Section 5 contains a brief summary of results and conclusions.

\section{OPTIMAL SEQUENTIALLY PLANNED TESTS}

In this section we adapt the results of
\cite{NovikovIMF} to the context of group sequential tests with variable group sizes. Following \cite{Schmitz}, we prefer the term ``sequentially planned test'' to ``group sequential test with variable group sizes''.

\subsection{Definitions and preliminaries}
We assume that a sequence of random variables $X_1,X_2,\dots,X_n,\dots$ will be available on the group-by-group basis for testing two simple hypotheses $H_0$ and $H_1$ about their distribution.

A group sequential test is based on a  family of rules  governing the process of sampling. Let us suppose the process of testing came to stage $i$ meaning that $i$ groups of observations have been taken, and let $n_1,n_2,\dots, n_i$ be the consecutive sizes of the groups taken.  Let the individual observations collected up to stage $i$ be $x_1,x_2,\dots x_{n_1+\dots+n_i}$.
Then the size of the next group to be taken at stage $i+1$ should be defined as a function $\psi_{n_1,\dots,n_i}=\psi_{{n_1,\dots,n_i}}(x_1,x_2,\dots x_{n_1+\dots+n_i})$ taking values in a set $G\cup\{0\} $, where $G\subset \mathbb N$ is finite.  A positive value of $\psi_{n_1,\dots,n_i}$ is interpreted as the size of the group of observations to be taken at stage $i+1$. If $\psi_{n_1,\dots,n_i}=0$, this means no more observations will be taken (and the process should be stopped with the data observed in $i$ groups). The size of the first group is defined before any observation is taken and will be denoted as $\psi_{()}$. In this way, a sampling plan is defined as the set of functions $\{\psi_{()}, \{\psi_{(n_1,\dots,n_i)}\}_{ n_i\in G, i=1,2,\dots, }\}$. 

After the sampling process is stopped, the final decision is taken using another element of the group sequential test called decision function and denoted as 
$ \{\phi_{(n_1,\dots,n_i)}\}_{ n_i\in G, i=1,2,\dots, }$. Any $\phi_{(n_1,\dots,n_i)}$ takes one of the values 0 or 1, meaning accepting the respective hypothesis $H_0$ or $H_1$. All the functions $\psi_{(n_1,\dots,n_i)}$ and $\phi_{(n_1,\dots,n_i)}$ are assumed to be measurable for all $(n_1,\dots,n_i)$ and all  $i=1,2,\dots$.

Let us denote $\langle \psi,\phi\rangle$ the sequentially planned test with the sampling rule $\{\psi_{()}, \{\psi_{(n_1,\dots,n_i)}\}_{ n_i\in G, i=1,2,\dots, }\}$ and the decision rule $ \{\phi_{(n_1,\dots,n_i)}\}_{ n_i\in G, i=1,2,\dots, }$.

%
%
%
%


A classical sequential test corresponds to $\psi_{(n_1,\dots,n_i)}=1$ or 0 (one observation at a time is taken, if any) for all $n_1,\dots n_i$, $i=1,2,\dots$. 

 Let us define $\nu_1^\psi=\psi_{()}$ and, recurrently over $i=1,2,\dots$,
 $\nu_{i+1}^\psi=\psi_{(\nu_1^\psi,\nu_2^\psi,\dots,\nu_i^\psi)}(X_1,\dots,X_{\nu_1^\psi+\nu_2^\psi+\dots+\nu_i^\psi})$

Let us define the following events:
$S_i^\psi=\{\nu_1^\psi>0,\dots, \nu_{i}^\psi>0, \nu_{i+1}^\psi=0\}$ (stop at stage $i$), and $C_i^\psi=\{\nu_1^\psi>0,\dots, \nu_{i}^\psi>0\}$ (stop at or after stage $i$).

Suppose there is some cost, say $c(n)$, we should pay for any group of $n$ items to be observed (for obvious reasons, we can assume that $c(n)>0$ for all $n\in G$; another natural assumption is that $c(n)$ is a strictly increasing function of $n$). Then the average sampling cost (ASC), under $H_j$,  of carrying out a test based on sampling plan $\nu$
is \begin{eqnarray}\label{28s_1}\mbox{ASC}_j(\psi)
&=&c(\nu_1^\psi)+\sum_{i=2}^\infty\mbox{E}_jc(\nu_i^\psi)I_{C_i^\psi}\nonumber
,\end{eqnarray}
 where E$_j$ is the symbol of the mathematical expectation calculated under hypothesis $H_j$, $j=0,1$.


Error probabilities of the first and the second kind are defined, respectively, as
\begin{equation*}
\alpha(\psi,\phi)= \sum_{i=1}^\infty\mbox{P}_0(S_i^\psi\cap \{\phi_{(\nu_1^\psi,\dots\nu_i^\psi)}=1\})
\end{equation*}
\begin{equation*}
\beta(\psi,\phi)= \sum_{i=1}^\infty\mbox{P}_0(S_i^\psi\cap \{\phi_{(\nu_1^\psi,\dots\nu_i^\psi)}=0\})
\end{equation*}

The usual context for hypothesis testing is  to minimise the average experimental cost under the restriction that
\begin{equation}\label{28_10}
\alpha(\psi,\phi)\leq \alpha \quad\mbox{and}\quad\beta(\psi,\phi)\leq\beta,
\end{equation}
where $\alpha$ and $\beta$ some numbers between 0 and 1.

In this paper, we want to minimise a weighted average
sampling cost
\begin{equation}\label{28s_12}
\mbox{ASC}_\gamma(\psi)=(1-\gamma) \mbox {ASC}_0(\psi)+\gamma\mbox {ASC}_1(\psi)\end{equation}
under condition \eqref{28_10}, where $\gamma\in[0,1]$ is a given fixed number. The value of $\gamma$ represents the grade of importance we attribute to ASC$_1$ in comparison with ASC$_0$  when designing the optimal test. The extreme values 0 or 1 correspond to minimisation of ASC under $H_0$ and $H_1$, respectively, regardless of  the value the average cost may have under the other hypothesis. It is most desirable that {\em one} test satisfying \eqref{28_10} minimise {\em both} average sampling costs \citep[just like the classical SPRT in the one-per-group case does, see][]{waldwolfowitz}. Unfortunately, there is no known result in the literature that could guarantee this property (which may be called Wald-Wolfowitz optimality) for the sequentially planned tests. Anyway, if a test  with error probabilities  $\alpha $ and $\beta$ could ever be found that minimises both $ASC_0$ and $ASC_1$ among all the tests subject to \eqref{28_10},  it should also minimise ASC$_\gamma$, whatever $\gamma\in[0,1]$.

It is easy to see that the problem of minimisation of \eqref{28s_12} under restrictions \eqref{28_10} reduces to the problem of minimisation of
\begin{equation}\label{28s_13}
\mbox{ASC}_\gamma(\psi)+\lambda_0\alpha(\psi,\phi)+\lambda_1\beta(\psi,\phi)
\end{equation}
with some non-negative $\lambda_0,\lambda_1$ \citep[see][Section 2]{NovikovIMF}. Essentially, this is a straightforward application of the Lagrange method to a problem of constrained minimisation. From this point of view, \eqref{28s_13} is interpreted as a Lagrangian function with constant multipliers $\lambda_0,\lambda_1$. The multipliers should be used to guarantee that for the test minimising \eqref{28s_13} equalities in \eqref{28_10} are attained.

On the other hand, if $0<\gamma <1$ then \eqref{28s_13} can be seen as Bayesian risk  \citep [cf. (2.1) in ][]{Ehrenfeld} in the Bayes formulation. In this case $\gamma$ can be interpreted as an a priori probability of hypothesis $H_1$, and $\lambda_0, \lambda_1 $ as some   characteristics related to the loss  due to incorrect decisions.

In a very usual way, it can be shown that there is a unique form for the decision function to be used in \eqref{28s_13} for its minimisation.

We need some additional notation for this. Let $f_j^n=f_j^n(x_1, \dots, x_n)$ be the Radon-Nikodym density of the distribution of $(X_1, X_2, \dots, X_n)$ under $H_j$, $j=0,1$, with respect to a product-measure $\mu^n$ ($n$ times $\mu$ by itself),  and let $f_\gamma^n=(1-\gamma)f_0^n+\gamma f_1^n$.

For any set of group sizes $\vec n_i=(n_1,\dots,n_i)$ let us denote $f_\gamma^{\vec n_i}=f_\gamma^{n_1,\dots,n_i}=f_\gamma^{n}=f_\gamma^{n}(x_1,\dots,x_n)$ where $n=n_1+n_2+\dots+n_i$, $i=1,2,\dots$
 Let us also define $\mu^{\vec n_i}=\mu^{n_1,\dots,n_i}$ as the product-measure $\mu^{n_1}\times\dots \times\mu^{n_i}$.

Then for any given sampling plan $\psi$, \eqref{28s_13} is minimised by the decision function $\phi$ defined as
\begin{equation}\label{1o.1}
\phi_{n}=I_{\{\lambda_0f_0^n\leq \lambda_1f_1^n\}}
\end{equation}
for any $n=(n_1,\dots,n_i)$, $i=1,2,\dots$

The corresponding minimum value of \eqref{28s_13} is 
\begin{equation*}
L(\psi)=\sum_{n=1}^\infty\int I_{S_n^\psi}\left((c(\nu_1^\psi)+ \dots+c(\nu_n^\psi))f_\gamma^{\vec \nu_n^\psi}+\min\{\lambda_0f_0^{\vec\nu_n^\psi},\lambda_1f_1^{\vec\nu_n^\psi}\}\right)d\mu^{\vec\nu_n^\psi},
\end{equation*}
where $\vec\nu_n^\psi=(\nu_1^\psi,\dots,\nu_n^\psi)$.
 The proof is along the lines of the proof of Theorem 3.1 in \cite{NovikovIMF}.

\subsection{Optimal truncated plans}

Let $\mathcal F^K$ be the set of sampling plans $\psi$ taking at most $K$  groups (such that $\nu_{K+1}^\psi\equiv 0$). 
For $\psi\in \mathcal F^K$, let us denote
\begin{equation*}
L_K(\psi)=\sum_{n=1}^{K}\int I_{S_n^\nu}\left((c(\nu_1^{\psi})+ \dots+c(\nu_n^{\psi}))f_\gamma^{\vec\nu_n}+\min\{\lambda_0f_0^{\vec\nu_n},\lambda_1f_1^{\vec\nu_n}\}\right)d\mu^{\vec\nu_n}.
\end{equation*}

Starting from
$$
V_{\vec n_K}^K=\min\{\lambda_0f_0^{\vec n_K},\lambda_1f_1^{\vec n_K}\}
$$
define  recursively for $i=K-1,K-2,\dots, 1$
\begin{equation*}
 V_{\vec n_i}^K=\min\{\lambda_0f_0^{\vec n_i},\lambda_1f_1^{\vec n_i},\min_m\{c(m)f_\gamma^{\vec n_i}+\int V_{\vec n_i,m}^Kd\mu^{m}\}\}
\end{equation*}
Then for any sampling plan $\psi\in \mathcal F^K$
\begin{equation}\label{30s.1}
 L_K(\psi)\geq \min_m \{c(m)+\int V_m^K d\mu^{m}\}
\end{equation}
There is an equality in \eqref{30s.1} if a sampling plan $\psi\in \mathcal F^K$ is such that   for  $i=1,2,\dots,K-1$
\begin{equation*}\label{30s_4}
\psi_{\vec n_i}=\begin{cases}0 \quad\mbox{if}\quad\min\{\lambda_0f_0^{\vec n_i},\lambda_1f_1^{\vec n_i}\}\leq\underset{m}{\text{min}}\{c(m)f_\gamma^{\vec n_i}+\int V_{\vec n_i,m}^Kd\mu^{m}\},\;\text{}\cr
\underset{m}{\mbox{argmin}}\{c(m)f_\gamma^{\vec n_i}+\int V_{\vec n_i,m}^Kd\mu^{m}\}, \text{ otherwise,}\end{cases}
\end{equation*}
and $\psi_{()}=\underset{m}{\mbox{argmin}}\{c(m)+\int V_{m}^Kd\mu^{m}\}$.

It follows from \eqref{30s.1} that this is an optimal sampling plan minimizing $L_K(\psi)$ in $\mathcal F^K$.

This result can be obtained in essentially the same way as Corollary 4.4 in \cite{NovikovIMF}.

\subsection{Optimal non-truncated plans}
The treatment of the general case is  essentially the same as in Section 5 in \cite{NovikovIMF}.

First, it can easily be shown that $V_{\vec n_i}^{K+1}\leq V_{\vec n_i}^K$ for any fixed $\vec n_i=(n_1,\dots , n_i)$, so there exists
$$
V_{\vec n_i}=\lim_{K\to\infty}V_{\vec n_i}^K.
$$
Furthermore, $L_K(\psi)\to L(\psi)$ as $K\to \infty$ for all $\psi \in\mathcal F$.  

Therefore, it follows from \eqref{30s.1} that for all $\psi \in\mathcal F$
\begin{equation}\label{30s.3}
 L(\psi)\geq \min_m \{c(m)+\int V_m d\mu^{m}\}.
\end{equation}
And similarly to the proof of Theorem 5.5 in \cite{NovikovIMF} it is shown that
there is an equality in \eqref{30s.3} if a sampling plan $\psi $ is such that
for  $i=1,2,\dots$
\begin{equation}\label{30s.41}
\psi_{\vec n_{i}}=\begin{cases}0 \quad\mbox{if}\quad\min\{\lambda_0f_0^{\vec n_{i}},\lambda_1f_1^{\vec n_{i}}\}\leq\underset{m}{\text{min}}\{c(m)f_\gamma^{\vec n_{i}}+\int V_{\vec n_{i},m}d\mu^{m}\},\;\text{}\cr
\underset{m}{\mbox{argmin}}\{c(m)f_\gamma^{\vec n_{i}}+\int V_{\vec n_{i},m}d\mu^{m}\}, \text{ otherwise,}\end{cases}\end {equation}
and $\psi_{()}=\underset{m}{\mbox{argmin}}\{c(m)+\int V_{m}d\mu^{m}\}$.

Therefore, this is  a form of a sampling plan which minimises the Lagrangian function $L(\psi)$, over all $\psi \in \mathcal F$. All other optimal sampling plans can be obtained from this one by randomisation, similarly to \cite{NovikovIMF}. The randomisation can be applied in any case of equality between the respective elements in \eqref{30s.41}, including those participating in the argmin definition. Obviously, the randomisation is irrelevant for the  Bayesian set-up but may be useful in the conditional setting.

\section{THE I.I.D. CASE}

In the case of independent and identically distributed (i.i.d.) observations the constructions of Section 2 acquire a much simpler form.

Let $f_j(x)$ be the Radon-Nikodym derivative (with respect to $\mu$) of the distribution of $X_i$ under hypothesis $H_j$, and assume $X_i$ are all independent, $i=1,2,\dots$. 
Very naturally, it should be assumed that the hypothesized distributions are distinct:
$ \mu(x: f_0(x)\not=f_1(x))>0$.

\subsection{Optimal sequentially planned tests }

 Let us define  operator $\mathcal I_m$,  defined  for any bounded measurable non-negative function $U(z), z\geq 0$, as 
\begin{equation*}
\mathcal I_mU=\mathcal I_mU(z)=E_0\left(U\left(z\frac{f_1^m(X_1,\dots,X_m)}{f_0^m(X_1,\dots,X_m)}\right)\right), \; z\geq 0.
\end{equation*}
Let us denote $g(z)=g(z;\lambda_0,\lambda_1)=\min\{\lambda_0,\lambda_1z\}$ for $z\geq 0$.

Starting from \begin{equation}\label{30s_3}
\rho_0(z)=g (z)
\end{equation}
define recursively over $i=1,2,\dots$
\begin{equation}\label{1o_2}
\rho_{i}(z)=\min\{g(z),\min_{m}\left\{ c(m)(1+\gamma(z-1))+\mathcal I_m\rho_{i-1}(z)\right\}\}.
\end{equation}
It is easy to see, by induction, that 
\begin{equation*}
 V_{\vec n_i}^K=\rho_{K-i}(z_{\vec n_i})f_0^{\vec n_i}
\end{equation*}
where 
$
 z_{\vec n_i}={f_1^{\vec n_i}}/{f_0^{\vec n_i}}
$
is the likelihood ratio.
 
 Furthermore, the optimal sampling rule $\psi \in\mathcal F^K$ depends on $z={z_{\vec n_{i}}}$, $\psi_{\vec n_i}=\hat t_i(z_{\vec n_i})$, whatever be  $\vec n_i$ and $i=2,3, \dots, K$, and $\psi_{()}=\hat t_1(1)$,
where for $i=2,3,\dots,K$ and $z\geq 0$
\begin{equation}\label{30s_4s}
\hat t_i=\hat t_i(z)=\begin{cases}0 \quad\mbox{if}\quad g(z)\leq\underset{m}{\text{min}}\{c(m)(1+\gamma(z-1))+\mathcal I_m\rho_{K-i}(z)\},\;\text{}\cr
\underset{m}{\mbox{argmin}}\{c(m)(1+\gamma(z-1))+\mathcal I_m \rho_{K-i}(z)\}, \text{ otherwise,}\end{cases}
\end{equation}
and $\hat t_1 (1)=\underset{m}{\mbox{argmin}}\{c(m)+\mathcal I_m \rho_{K-1}(1)\}$.

In particular, we obtain here a characterisation of truncated Bayesian sequentially planned test (properly saying, of its sampling-plan part, but the decision function to apply with any sampling plan is universal and is defined in \eqref{1o.1}, so we obtain a complete sequentially planned test $\langle \psi,\phi \rangle$ which is Bayes-optimal). In fact, \eqref{1o.1} can also be expressed in terms of the likelihood ratio:
\begin{equation}\label{1o.1bis}
\phi_{\vec n_i}=I_{\{\lambda_0\leq \lambda_1z_{\vec n_i}\}}
\end{equation}
for any $n=(n_1,\dots,n_i)$, $i=1,2,\dots$

 In the same way, to obtain  optimal truncated sequentially planned tests in the conditional set-up we only need to satisfy \eqref{28_10} choosing appropriate Lagrangian multipliers $\lambda_0$ and $\lambda_1$.

At last, a concluding remark on the form of continuation regions of optimal truncated tests.
Using the concavity of functions $\rho_i$, $\mathcal I_m\rho_i$ and other involved, it is not difficult to see that the ``continuation region''  $\{z: \hat t_{i}(z)>0\}$ (see the first line of \eqref{30s_4s}) always has a form of an interval $(a_i, b_i)$ (if not empty). 
We used this technique in \cite{NovikovPopoca} for a related problem, when the group sizes are independent of  the observations. In addition, it can be shown that $a_1\leq a_2\leq\dots\leq a_{i}$ and $b_1\geq b_2\geq\dots\geq b_{i}$, where $(a_i,b_i)$ is the last non-empty continuation interval.

The optimal non-truncated tests are obtained by passing to the limit as $K\to\infty$.
It is easy to see that $\rho_K(z)\geq\rho_{K+1}(z) $ for all $z\geq 0$ (see \eqref{1o_2}). So there exists $\rho(z)=\lim_{K\to\infty}\rho(z)$, $z\geq 0$.
It is a concave non-decreasing function with $\rho(0)=0$.

Passing to the limit in \eqref{1o_2} as $i\to\infty$, we obtain 
\begin{equation*}\label{1o_2a}
\rho(z)=\min\{g(z),\min_{m}\left\{ c(m)(1+\gamma(z-1))+\mathcal I_m\rho(z)\right\}\}.
\end{equation*}

Now, passing to the limit in \eqref{30s_4s}, as $K\to\infty$ we see that the {\em optimal} sampling plan  $\psi \in\mathcal F$ depends on $z={z_{\vec n_{i}}}$, in such a way that  $\psi_{\vec n_i}=\hat t(z_{\vec n_i})$, for any $\vec n_i$ and $i=1,2, \dots, $ and $\psi_{()}=\hat t(1)$,
where for any $z\geq 0$
\begin{equation}\label{30s_4sa}
\hat t=\hat t(z)=\begin{cases}0 \quad\mbox{if}\quad g(z)\leq\underset{m}{\text{min}}\{c(m)(1+\gamma(z-1))+\mathcal I_m\rho(z)\},\;\text{}\cr
\underset{m}{\mbox{argmin}}\{c(m)(1+\gamma(z-1))+\mathcal I_m \rho(z)\}, \text{ otherwise,}\end{cases}
\end{equation}

Once again, the continuation region $\{\hat t(z)>0\}=(a,b)$ is an interval  with some $0<a<b<\infty$ (if not empty).

The resulting sequentially planned test $\langle \psi,\phi\rangle$ is a particular case of  Secuentially Planned Probability Ratio Tests (SPPRTs) \cite[see (3.5) in ][]{Schmitz}, with  $k_1=a$, $k_2=b$.

In fact, the formal definition of SPPRT in \cite{Schmitz} requires that $k_1<1<k_2$. Our construction \eqref{30s_4sa} may result in $1\not\in (a,b)$, but it is still  optimal in the class of sequentially planned tests we consider (those taking at least one group of observations). 
\subsection{Performance characteristics}

In this part, we obtain formulas for calculating error probabilities, average sampling cost and some  related probabilities, for  truncated sequentially planned  tests.

 Let  $\psi \in\mathcal F^K$ be a sampling rule depending on $z={z_{\vec n_{i}}}$, in such  a way that  $\psi_{\vec n_i}=\hat t_i(z_{\vec n_i})$, whatever be  $\vec n_i$ and $i=2,3, \dots, K-1$, and $\psi_{()}=\hat t_1(1)$, and let $\phi$ be any decision rule such that $\phi_{\vec n_i}=\delta(z_{\vec z_i})$ for all $\vec n_i$ and $i=1,2,\dots, K$.



Let us denote $A_i^K(=A_i^K(\psi,\phi))$  the event meaning 
 ``H$_0$ is accepted  at stage $i$ or thereafter" (in accordance with the rules of the test $\langle \psi,\phi\rangle$).

\begin{proposition}\label{P2} 
Let \begin{equation}\label{25a_1}d_K^K(z)=I_{\{\delta(z)=0\}}(z),\; z\geq 0,\end{equation} and, recursively over $i=K-1,K-2,\dots,1$, 
\begin{equation}\label{25a_2}d^K_{i}(z)=I_{\{\hat t_{i}=0\}}(z)I_{\{\delta=0\}}(z)+\sum_{m>0}I_{\{\hat t_{i}=m\}}(z)\text{E}d_{i+1}^K(zZ_{m}),\;z\geq 0.\end{equation}

Then for any  $1\leq i\leq K$
\begin{equation}\label{22e_1} d^K_i(Z_{\vec \nu_i^\psi})=\text{P}({A_i^K}| Z_{\vec \nu_i^\psi} ),\end{equation}  
 in particular,
 \begin{equation}\label{6o.2}
 \text{P}(A_1 ^K)=\text{E} d^K_1(Z_{ \nu_1^\psi}).
 \end{equation}
\end{proposition}
\begin{remark}The distribution for use in  Proposition \ref{P2} (in \eqref{25a_2}, \eqref{22e_1} and \eqref{6o.2}) is primarily ment to be the one hypothesized under  $H_0$ or $H_1$,  but in principle it  may be any particular distribution preserving  the i.i.d. structure of the observations $X_1, X_2,\dots$.

As a particular case,  for $\psi\in \mathcal F^K$
\begin{equation*}
 \alpha(\psi,\phi)=1-\text{P}_0(A_1^K) \quad\text{and}\quad \beta(\psi,\phi)=\text{P}_1(A_1^K),
\end{equation*}
so the error probabilities can be calculated using $H_0$ and $H_1$, respectively, in the  evaluation of $d_i^K$ in Proposition \ref{P2}.

In the same way, the operating caracteristic of the truncated SPPRT in this case can be calculated just using any third distribution of $X_i$ in    Proposition \ref{P2}.
\end{remark}

{\bf Proof} of Proposition \ref{P2}. By induction over $i=K,K-1,\dots,1$.

For $i=K$, it follows from \eqref{25a_1} that $
d_K^K(Z_{\vec \nu_K^\psi})=I_{\{\delta=0\}}(Z_{\vec\nu_K^\psi})=\text{P} \{A^K_K|Z_{\vec \nu_K^\psi}\}$
(this latter equality follows from the definition of the decision function).

Let us suppose now that \eqref{22e_1} holds for some $2\leq i\leq K$.

Then 
\begin{eqnarray*}
d^K_{i-1}(Z_{\vec \nu^\psi_{i-1}})
&=&I_{\{\hat t_{i-1}=0,\delta=0\}}(Z_{\vec \nu_{i-1}^\psi})+\quad\sum_{m>0}I_{\{\hat t_{i-1}=m\}}(Z_{\vec\nu_{i-1}^\psi})\text{E}\{d^K_{i}(Z_{\vec\nu_{i-1},m}^\psi)|Z_{\vec\nu_{i-1}^\psi}\}
\\
&=&I_{\{\hat t_{i-1}=0,\delta=0\}}(Z_{\vec\nu_{i-1}^\psi})+\quad\sum_{m>0}\text{E}\{I_{\{\hat t_{i-1}=m\}}(Z_{\vec\nu_{i-1}^\psi})\text{E} \{I_{A_i^K}|Z_{\vec\nu_{i-1}^\psi,\nu_i^\psi}\}|Z_{\vec\nu_{i-1}^\psi}\}\\
&=&I_{\{\hat t_{i-1}=0,\delta=0\}}(Z_{\vec\nu_{i-1}^\psi})+\sum_{m>0}I_{\{\hat t_{i-1}=m\}}(Z_{\vec\nu_{i-1}^\psi})\text{E}\{I_{A_i^K}|Z_{\vec\nu_{i-1}^\psi}\}\\
&=&\text{E}\{I_{\{\hat t_{i-1}=0,\delta=0\}}(Z_{\vec\nu_{i-1}}^\psi)+\sum_{m>0}I_{\{\hat t_{i-1}=m\}}(Z_{\vec\nu_{i-1}^\psi})I_{A_i^K}|Z_{\vec\nu_{i-1}^\psi}\}\\
&=&\text{E}\{I_{\{\nu_{i-1}^\psi=0\}}I_{\{\delta(Z_{\vec\nu_{i-1}^\psi})=0\}}+I_{\{\nu_{i-1}^\psi>0\}}I_{A_i^K}|Z_{\vec\nu_{i-1}^\psi}\}
=\text{P}\{A_{i-1}^K|Z_{\vec\nu^\psi_{i-1}}\}\\
\end{eqnarray*}
$\Box$

 Analogously, the average sampling cost can be evaluated as follows.
 
 \begin{proposition}\label{P3}
 Let  \begin{equation*}\label{23e_4}
                        l_{K}^K(z)=0
                       \end{equation*}
  and, recursively over $i=K-1,K-2,\dots,2$, 
 \begin{equation}\label{23e_5}
 l_{i}^{K}(z)=\sum_{m>0}I_{\{\hat t_i=m\}}(z)(c(m)+\text{E} l_{i+1}^{ K }(zZ_{m})),z\geq 0.
 \end{equation}

 Then for all $1\leq i\leq K-1$
 \begin{equation}\label{23e_3}
 l_{i}^K(Z_{\vec\nu_{i}^\psi})=\text{E}\{(c(\nu_{i+1}^\psi)I_{\{\nu_{i}^\psi>0\}}+\dots+c(\nu_{K}^\psi)I_{\{\nu_{i+1}^\psi\dots \nu_{K-1}^\psi>0\}}\} |Z_{\vec\nu^\psi_{i}}\}
 \end{equation}

 In particular, 
 \begin{equation}\label{4o.1}
 \text{E}(c(\nu_2^\psi)I_{\{\nu_{1}^\psi>0\}}+\dots+c(\nu_{K}^\psi)I_{\{\nu_{2}^\psi\dots \nu_{K-1}^\psi>0\}})=\text{E}l_{1}^K(Z_{\nu^\psi_1})
  \end{equation}
\end{proposition}
\begin{remark}
 It is easy to see that \eqref{4o.1} is equivalent to
\begin{equation}\label{4o.1ext}
 \text{E}(c(\nu^\psi_1)+c(\nu^\psi_2)I_{C^\psi_2}\dots+c(\nu^\psi_{K})I_{C^\psi_{K}})=c(\nu^\psi_1)+\text{E}l_{1}^K(Z_{\nu^\psi_1})
  \end{equation}
  (this is because $\nu^\psi_1=\text{const}>0$), and that the left-hand side of \eqref{4o.1ext}
  is the average sampling cost of using plan $\psi\in \mathcal F^K$.
  
  Again, arbitrary  distribution can be  used within Proposition \ref{P3} (and consequently in \eqref{4o.1ext}). 
In particular, using the distribution under $H_0$  and under $H_1$  when evaluating \eqref{23e_5} we obtain  from \eqref{4o.1ext}, respectively,  the average  sampling cost $ASC_0(\psi)$ and $ASC_1(\psi)$ and thus the value of 
\begin{equation}\label{5o.1}
ASN_\gamma(\psi)
=(1-\gamma)ASC_0(\psi)+\gamma ASC_1(\psi))
\end{equation}
for any $\psi\in\mathcal F^K$.

  \end{remark}

{\bf Proof} of Proposition \ref{P3}.

By definition, 
  \begin{equation*}\label{23e_3bis}
 l_{K}^K(Z_{\nu^\psi_1,\dots,\nu^\psi_{K-1}})=c(\nu^\psi_K)I_{\{\nu^\psi_K>0\}}=\text{E}\{c(\nu^\psi_K)I_{\{\nu^\psi_K>0\}}|Z_{\nu^\psi_1,\dots,\nu^\psi_{K-1}}\}.
 \end{equation*}
Let us suppose now that \eqref{23e_3} is satisfied for some $2\leq i\leq K-1$. Then, by definition,
 \begin{equation*}
 l_{i-1}^{K}(z)=\sum_{m>0}I_{\{\hat t_{i-1}=m\}}(z)(c(m)+\text{E} \{l_{i}^{ K }(zZ_m)\}) 
 \end{equation*}
 and
$$
l_{i-1}^{K}(Z_{\vec \nu^\psi_{i-1}})=I_{\{\nu_{i-1}^\psi>0\}}(Z_{\vec\nu_{i-1}^\psi})(c(\nu^\psi_{i})+\text{E} \{l_{i}^{ K }(Z_{\vec\nu^\psi_{i-1}}Z_{ \nu^\psi_{i}}|Z_{\vec\nu^\psi_{i-1}}\}))
$$
$$
=I_{\{\nu^\psi_{i-1}>0\}}(c(\nu^\psi_{i})
+\text{E} \{
\text{E}(c(\nu^\psi_{i+1})I_{\{\nu_{i}^\psi>0\}}+\dots+c(\nu^\psi_K)I_{\{\nu^\psi_{i+1}>0,\dots,\nu^\psi_{K-1}>0\}})|Z_{\vec\nu^\psi_{i
}}\}|Z_{\vec\nu^\psi_{i-1}}
\}$$
$$
=\text{E}\{I_{\{\nu^\psi_{i-1}>0\}}c(\nu^\psi_{i})+ 
c(\nu^\psi_{i+1})I_{\{\nu^\psi_{i-1}>0,\nu^\psi_i>0\}}+\dots+c(\nu^\psi_K)I_{\{\nu^\psi_{i-1}>0,\nu^\psi_i>0,\dots,\nu^\psi_{K-1}>0\}}|Z_{\vec \nu^\psi_{i-1}}
\},
$$
which proves \eqref{23e_3}  also for $i-1$.
$\Box$

  \begin{remark}
For  non-truncated SPPRTs $\langle \psi,\phi \rangle $, we  can also make use of 
Propositions \ref{P2} and \ref{P3}, by truncating the sampling plan $\psi$.

Let us define, for any $\psi\in \mathcal F$, $\psi^K$ as  the sampling plan $\psi$ redefined  it in such a way that
$\psi_{\vec n_{K}}\equiv 0$ for all $ \vec n_{K}$  (keeping intact all other components of $\psi$).  

Using the fact that $ASC_j(\psi^K)\to ASC_j(\psi)$, as $K\to\infty$, for  $j=0,1$, we obtain from \eqref{5o.1} a numerical approximation for $ASC_\gamma(\psi)$.

Also $\alpha(\psi,\phi)=\lim_{K\to\infty}(1-P_0(A_1^K))$ and $\beta(\psi,\phi)=\lim_{K\to\infty}P_1(A_1^K)$, so we obtain from Proposition \ref{P2} a numerical approximation for the  error probabilities $\alpha(\psi,\phi)$ and $\beta(\psi,\phi)$.
\end{remark}
\begin {remark}
Proposition \ref{P3} can also be used for calculating other sampling characteristics of the test. 

 For example, taking in Proposition \ref{P3} $c(m)=1$, for all $m$, one obtains, in place of the ASC, the average number of groups taken, E$T$.

Employing $c( m)=m$, for all $m$, provides the average number of observations taken, $\text{E}(\nu^\psi_1+\nu^\psi_2+\dots+\nu^\psi_T).$
\end{remark}
\section{NUMERICAL ALGORITHMS}
In this section we propose numerical algorithms for optimal design of sequentially planned tests and their performance evaluation.
\subsection{Optimal design and performance evaluation}
We propose a numerical method based on 
the optimal sampling  plan described in \eqref{30s_4s}.

The sampling plan is entirely based on the sequence of functions $\rho_0,\rho_1,\dots, \rho_{K-1}$ defined in \eqref{30s_3} and \eqref{1o_2} for $z\geq 0$. The idea of the  method is a numerical approximation of every $\rho_i$ on the continuation interval by a picewise-linear function based on a grid of $z$-values. So instead of functions $\rho_i$ we will work with functions $\tilde \rho_i$ defined as follows: let $\tilde \rho_0=g$, and define  recursively for $i=1,2,\dots$
\begin{equation*}\label{1o_2bis}
\tilde \rho_{i}(z)=\min\{g(z),\min_{m}\left\{ c(m)(1+\gamma(z-1))+\mathcal I_m\tilde \rho_{i-1}(z)\right\}\},
\end{equation*}
where $\tilde\rho_i$ is calculated by interpolation between the grid points on the continuation interval.

Formally, the proposed algorithm is as follows (applicable for $ K\geq 2$).
{\flushleft \em Numerical implementation of the optimal design (NIOD algorithm)}
\begin{description}
 \item 
[Step 1] Start from $n=1$
\item [Step 2] Find a minimum ($a_n$) and a maximum ($b_n$) value of $z$ 
for which
\begin{equation}\label{1o.10}
g(z)>\underset{m}{\text{ min}}
\{c(m)(1+\gamma(z-1))+\mathcal I_m \tilde \rho_{n-1}(z)\}
\end{equation}
If no such $z$ exist, declare Early Exit condition and Stop.
\item[ Step 3] For a grid $\{z_i\}$ of values on $[a_n,b_n]$ calculate and store the respective values $\{v_i\}$ of the function on the right in \eqref{1o.10}. 
Take note, for future use, that
$
\tilde\rho_n(z)
$
will be calculated as $g(z)$ for $z\not\in(a_n,b_n)$ and using an interpolation between the respective grid points for $z\in[a_n,b_n]$
\item [Step 4] Set $n=n+1$. If $n=K$ then Stop, else go to Step 2.
\end{description}

The Early  Exit condition means that the optimal testing procedure in fact is truncated at a lower level than $K$, because at step  $n$ there is no continuation interval, i.e., $\nu^\psi_n(z)\equiv 0$, $z\geq 0$. We can incorporate the Early Exit condition into the general scheme simply adjusting the truncation level by taking as $K$ the minimum of $K$ and $n$. In particular, it may happen that $n=1$, meaning that
 only one-stage sampling plans come into question in the hypothesis testing problem with given input parameters (for example, when the cost of data is too high).

In case the algorithm terminates with one stage, the first sample of size   $\nu^\psi_1=\underset{m}{\mbox{argmin}}\{c(m)+\mathcal I_m g(1)\}$ is taken, and the decision function \eqref{1o.1} with $n=\nu^\psi_1$ is applied.

After the algorithm stops with $K\geq 2$  (corrected for the Early Exit, if applicable), we obtain a way to calculate the functions  $\tilde \rho_1$, $\dots$, $ \tilde \rho_{K-1}$ and to apply them in definitions of optimal sampling plans in \eqref{30s_4s} (where $i=2,3,\dots, K$ and $z=z_{n_1,\dots n_{i}}$):
\begin{equation}\label{3o.1}
\tilde t_{i}=\tilde t_i(z)=\begin{cases}0 \quad\mbox{if}\quad z\not\in[a_{K-i+1},b_{K-i+1}],\cr
\underset{m}{\mbox{argmin}}\{c(m)(1+\gamma(z-1))+\mathcal I_m\tilde \rho_{K-i}(z)\}, \text{ otherwise,}\end{cases}
\end{equation}
and $\tilde t_1=\underset{m}{\mbox{argmin}}\{c(m)+\mathcal I_m \tilde\rho_{K-1}(1)\}$, $\nu^{\tilde \psi}_{K+1}\equiv 0$ (let us denote $\tilde\psi_{\vec n_i}=\tilde t_i(z_{\vec n_i})$ for any $\vec n_i$ and any $i$).

Respectively, we can use Proposition \ref{P2} for approximate evaluation of error probabilities by substituting $\tilde t_i$ for $t_i$ in \eqref{25a_2} for $i=K-1,\dots,1 $:
\begin{equation}\label{6o.1}\tilde d^K_{i}(z)=I_{\{\tilde t_{i+1}=0\}}(z)I_{\{  \delta_i=0\}}(z)+\sum_{m>0}I_{\{\tilde t_{i+1}=m\}}(z)\text{E}_1\tilde d_{i+1}^K(zZ_{m}),\;z\geq 0.\end{equation}
with $\tilde d_K^K\equiv d_K^K$, and finally from  \eqref{6o.2}
\begin{equation}\label{7o.1}
\beta(\tilde \psi,\phi)=\text{E}_1 \tilde d_1^K(Z_{ \nu^{\tilde \psi}_1})
\end{equation}
as an approximate value of $\beta(\psi,\phi)$.

Analogously, using E$_0$ instead of E$_1$ in \eqref{6o.1} for $i=K-1,\dots,1 $ we get an approximation
for $\alpha(\psi,\phi)$ in the form of  
\begin{equation}\label{7o.2}
\alpha(\tilde \psi,\phi)=1-\text{E}_0\tilde d_1^K(Z_{ \nu^{\tilde \psi}_1}).
\end{equation}

In the same way, substituting $\nu^{\tilde \psi}_i$ for $\nu^\psi_i$ (and $\tilde l_i^K$ for $l_i^K$) in Proposition \ref{P3} we obtain from \eqref{4o.1ext} an aproximation for the average sampling cost:
\begin{equation}\label{6o.5}
 \text{E}(c(\nu^{\tilde  \psi}_1)+\dots+c( \nu^{\tilde \psi}_K)I_{\{\nu^{\tilde \psi}_2>0, \nu^{\tilde \psi}_3>0,\dots, \nu^{\tilde  \psi}_{K-1}>0\}})=c( \nu^{\tilde  \psi}_1)+\text{E}\tilde l_{1}^K(Z_{ \nu^{\tilde \psi}
 _1}),
  \end{equation}
  whatever  the distribution of the i.i.d. observations  is used for calculations in Proposition \ref{P3}.

  All the computations involve some  values of the Lagrange multipliers $\lambda_0$ and $\lambda_1$ that should be determined in such a way that there are equalities in \eqref{28_10}.   In fact, for some $\alpha$ and $\beta$ this holds automatically, namely, for $\alpha=\alpha(\psi,\phi)$ and $\beta=\beta(\psi,\phi)$, where $\langle \psi,\phi\rangle$ is a test minimising \eqref{28s_13} for  some $\lambda_0, \lambda_1$. For other $\alpha$ and $\beta$,  there is no way to guarantee the existence of  $\lambda_0$ and $\lambda_1$ providing equalities in \eqref{28_10}, not even in the classical case of purely sequential tests.

  To conclude this subsection, let us summarize the algorithm of the optimal test evaluations. The overall procedure is quite straightforward.
  {\flushleft \em Optimal test evaluations (OTE algorithm)}
  \begin{description}
   \item [Input parameters:] hypothesis points $\theta_0$, $\theta_1$, Lagrangian multipliers $\lambda_0$, $\lambda_1$, grid size $h$, horizon $K$, the set of eligible group sizes $G$, cost function $c(m), m\in G$, weight parameter $\gamma$.
   \item [Step 1] Run the algorithm NIOD above in this subsection.
   \item [Step 2] Calculate the optimal sampling plan $\tilde t_i$ $i=1,\dots, K-1$ (see \eqref{30s_4s}).
  \item[Step 3] Calculate the error probabilities $\alpha(\tilde \psi,\phi)$ (see \eqref{7o.2}) and $\beta(\tilde \psi,\phi)$ (see \eqref{7o.1}).
  \item [Step 4] Calculate the average sampling cost: ASC$_0(\tilde \psi)$ and ASC$_1(\tilde \psi)$, using \eqref{6o.5}.
\item [Step 5] Calculate the Average Number of Observations and/or Average Number of Groups (optional, see Remark 3.4).
  \item[Output:] $\alpha(\tilde \psi,\phi)$ and  $\beta(\tilde \psi,\phi)$, ASC$_0(\tilde \psi)$ and ASC$_1(\tilde \psi)$ (optionally, the Average Number of Observations and/or the Average Number of Groups).
  \end{description}
 We implemented this algorithm for the problem of testing   hypothesis $\theta=\theta_0$ vs. $\theta=\theta_1$ about the success probability $\theta$ of a Bernoulli distribution. The program code in R programming language \citep{R} can be downloaded from a public GitHub repository at 
  {\tt\small https://github.com/HOBuKOB-MEX/SPPRT}.  There is an R function for each step of the above OTE algorithm in the program implementation. The documentation is provided in the repository.

With the program code at hand, it is easy to make the output plan  $\langle \tilde \psi,\phi\rangle$ satisfy  restrictions \eqref{28_10} by varying the input parameters $\lambda_0$ and $\lambda_1$ in a series of  trial-and-error iterations.  
The following empirical fact is very helpful for doing the work.  The main effect of $\lambda_0$ is on $\alpha(\tilde \psi,\phi)$: larger values make   $\alpha(\tilde \psi,\phi)$ smaller, leaving $\beta(\tilde \psi,\phi)$ largely unaffected; similarly, increasing $\lambda_1$
mainly affects    $\beta(\tilde \psi,\phi)$ making it smaller, with no significant change in $\alpha(\tilde \psi,\phi)$.
  

  \subsection{Numerical examples}
 
In this subsection, we apply the program code for two practical examples of the optimal sequentially planned tests.

\subsubsection{Numerical comparison of sampling plans when testing for majority}
 \cite{SchmegnerBaron} provided a technique for numerical evaluation of the SPPRTs based on general results for random walks, in the particular case when the log-likelihood takes its values on a lattice. The results are applicable, in particular, for testing a hypothesis $H_0: \theta = \theta_0$ vs.  $H_1:\theta=\theta_1=1-\theta_0$ for the success probability $\theta$ of a Bernoulli distribution.  Using the SPPRT with  various sampling plans known from the literature  (see the  details of the plans ibid.), \cite{SchmegnerBaron} evaluated the SPPRT for testing $H_0: \theta=0.52$ vs. $H_1:\theta =0.48$ in a series of scenarios, searching for the most inexpensive sampling plan in a particular practical context,  with respect to the cost function $c(m)=c_0+cm$, where $c=10$ (cost per observation) and $c_0=1000$ (cost per group). All the evaluated sampling plans were based on a continuation interval which guaranteed that the error probabilities of the first and second kind were at most   $\alpha=\beta=0.05$. The best value of ASC found was 18254, with an average number of groups of E$T=9.3$ 
and an average total number of observations of E$M=892$ \citep[see][Example 4.6]{SchmegnerBaron}.

   We present here the results of numerical evaluations corresponding to our  optimal test evaluations algorithm in Subsection 4.1, in exactly the same context, to be able to compare the performance of our plan with those in \cite{SchmegnerBaron}.

For our implementation of the algorithms of  Subsection 4.1, we used, in each continuation interval, a uniform grid formed by equidistant points on the logarithmic scale of $z$ with step $h=0.1$. We used our truncated SPPRT \eqref{3o.1} with $K=15$ (maximum number of groups to be observed). Also we employed as the set $G$ of possible group sizes the set $G=\{10i+10\}_{i=0,\dots,59}$. We used  ASC$_\gamma$ with $\gamma=0.5$ as a criterion of minimisation.  The Lagrange multipliers $\lambda_0=\lambda_1=44$ were chosen in such a way that  
  $\alpha(\tilde \psi,\phi)=0.05$ and  $\beta(\tilde \psi,\phi)=0.05$.
  
 Evaluating the  characteristics of the proposed SPPRT according to \eqref{7o.1} - \eqref{6o.5} we obtained  $\text{ASC}_0=\text{ASC}_1=11510$, with   the average number of groups E$T=2.07$ and the average number of observations E$M=944$.
 Thus,  our method provides nearly 1.6 times lower  sampling cost, in comparison with the best plan found in \cite{SchmegnerBaron}.  Taking into account that there are a number of  ways to improve the numerical characteristics, namely, by 1) choosing a higher truncation level $K$, 2) making the grid size $h$ smaller, 3) making the set $G$ of eligible group sizes ``denser'', 4) adjusting the  criterion of minimization by varying $\gamma$ as required by the practical context, -- taking this into account, the real efficiency of the proposed method can turn even higher.
 
In Figure 1, the set of continuation intervals at each of the fourteen sequentially planned steps is presented. We know that theoretically the continuation interval gets closer to the  
one corresponding to the optimal  non-truncated SPPRT, as $K\to\infty$.  It appears that in this example the convergence is so  fast that the interval reaches its limit after as few as some 4 steps (remember that the first interval found  comes  last). 

In  Figure 2, one can see the ``nearly optimal'' sampling plan $\tilde t(z)$ calculated as a NIOD approxination to \eqref{30s_4sa}.
Again, because  $ \psi^K(z)$ converges to  the optimal sampling plan $\psi$ as $K\to\infty$,  we may expect that this is approximately the sampling plan the optimal non-truncated SPPRT will use in each step.  
 \begin{figure}[!t]
   \includegraphics[scale=0.7]{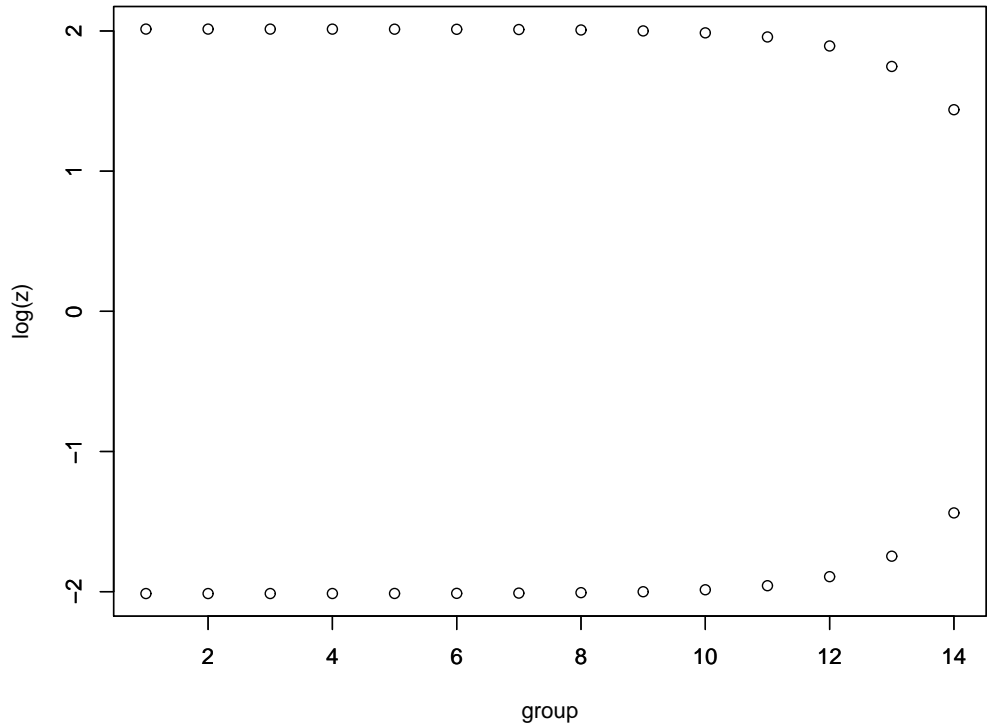}  
\caption{Nearly optimal truncated ($K=15$) stopping plan}
 \end{figure}
  \begin{figure}[!t]
    \includegraphics[scale=0.7]{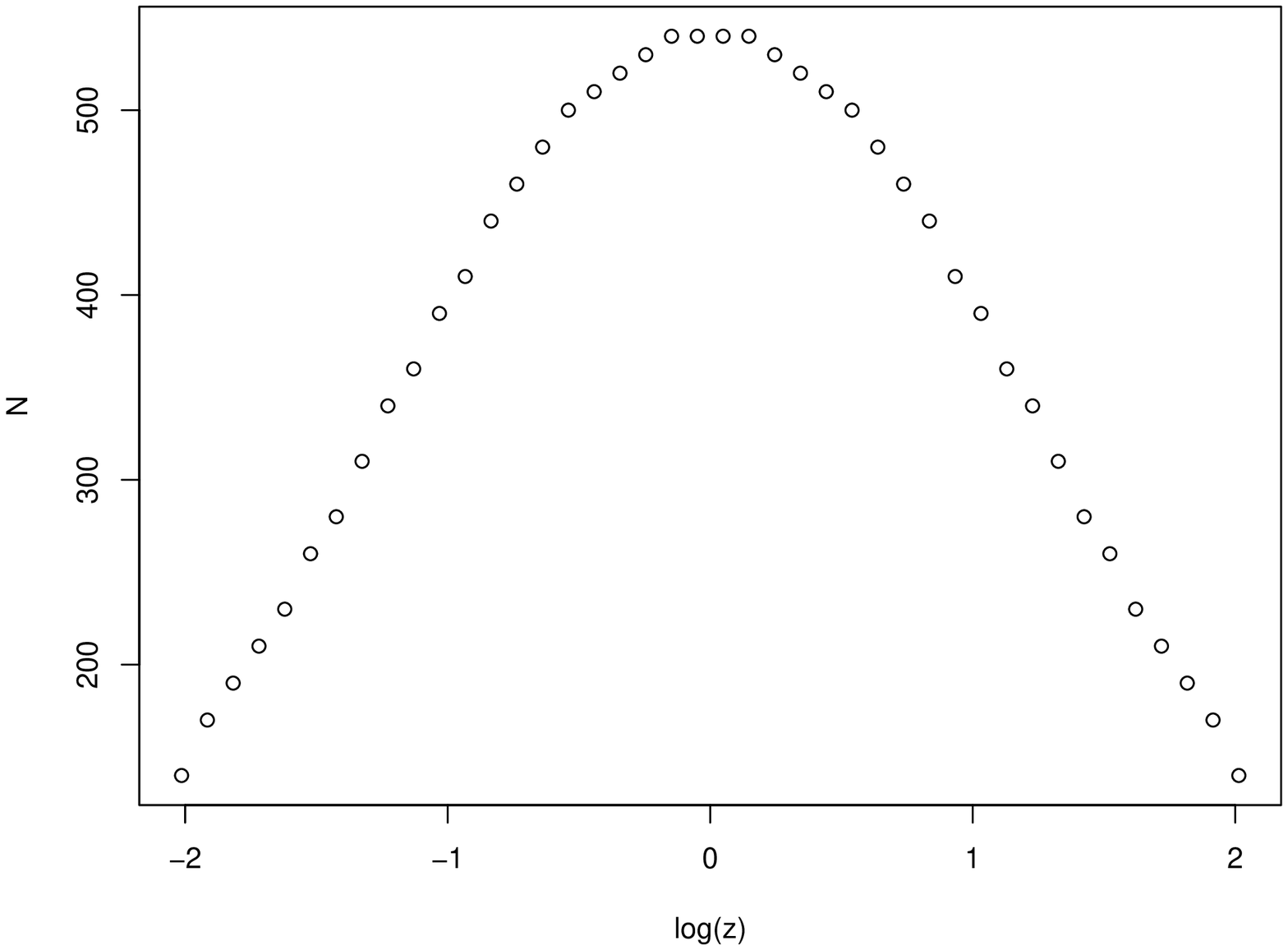}
\caption{Nearly optimal sampling plan $\tilde t(z)$}
 \end{figure}
  
 Taking into account that the gain from using our method, with respect to other known methods, is comparable to the gain  the classical SPRT provides with respect to one-sample (fixed sample size, FSS) test, we would like to examine the efficiency of our method with respect to the one-sample  test, in  various scenarios.
 As a reference, we want to use, for given $\theta_0$ and $\theta_1$, the average sampling cost of the one-step test with a minimum sample size $n(\alpha,\beta)$ that provides  error probabilities not exceeding $\alpha$ and $\beta$, respectively. For the Bernoulli model, $n(\alpha,\beta)$ can be calculated using the {\tt NP} function from the GitHub repository \cite{GitHub}.

 According to the definition of average sampling cost, the one-sample test has an ASC equal to  ASC$_{FSS}=c_0+cn(\alpha,\beta)$,
 which will be compared with ASC of the SPPRT we proposed.

 It is interesting to note that for $\alpha=\beta=0.05$ and $\theta_0=0.52$ and $\theta_1=0.48$ used in this example  the FSS $n(\alpha,\beta)$ is equal to $1691$, giving the average sampling cost of ASC$_{FSS}=17910$ for the one-sample plan, which outperforms all the SPPRTs examined in \cite{SchmegnerBaron}.
 
 To compare the performance of our ``nearly optimal'' SPPRT with that of  the one-sample test we ran our program for a series of $\lambda_0$ and $\lambda_1$ between 3 and 6.3 on the scale of natural logarithms, with a total of 9$\times$ 9 points. For each one, we calculated the corresponding $\alpha$, $\beta$ and ASC$_0$ and ASC$_1$. The relative efficiency was calculated  as $R_j=\text{ASC}_{FSS}/\text{ASC}_j$ under hypothesis $H_j$, $j=0,1$.
 
 To get a compact visual representation of the results, we fit a local polynomial regression model (LOESS)\footnote{https://www.rdocumentation.org/packages/stats/versions/3.6.2/topics/loess} to the data obtained,  to represent the relationship between  the relative efficiency and   $\alpha$ and $\beta$. We use decimal logarithms of $\alpha$ and $\beta$ as independent variables, and $R_j$ as response. The result of the model fitting for $R_0$ is shown in Figure 3. The graph of $R_1$ is perfectly symmetric with respect to the diagonal $\alpha =\beta$ and is not shown.

We see from Figure 3 that the  maximum of relative efficiency $R_0$ is attained in the asymmetric case when 
 $\alpha$ is small and  $\beta$ is relatively large, and is about 2.5. In the vicinity of the diagonal   the maximum efficiency of approx. 2.1 is reached for small $\alpha\approx\beta$ with a clear tendency of increasing as $\alpha,\beta\to 0$. The minimum efficiency is about 1.3 and is attained whenever $\alpha $ is relatively large.

 \begin{figure}[!t]
  \includegraphics[scale=0.7]{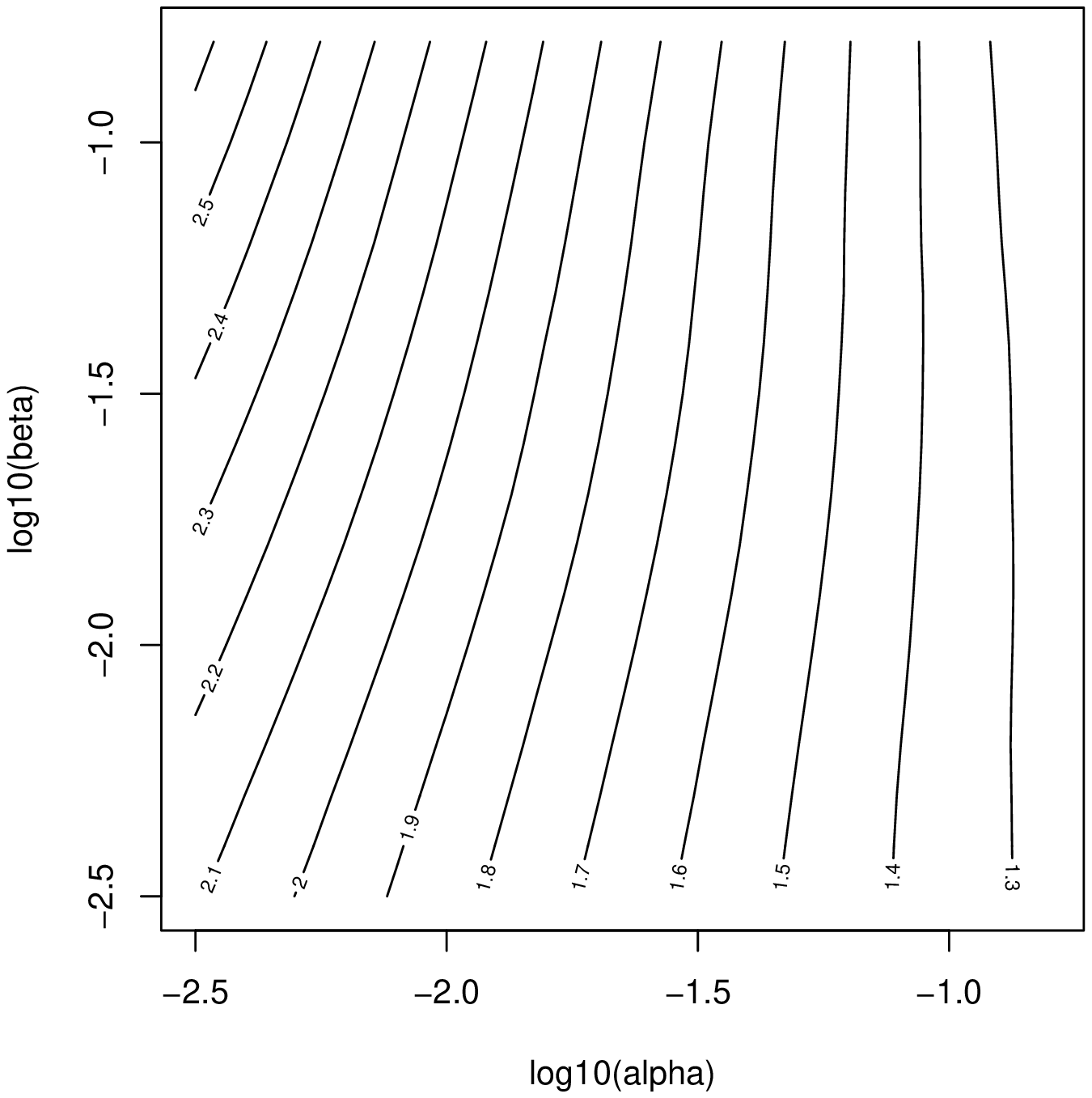}
\caption{Relative efficiency $R_0$ under $H_0$}
 \end{figure}

 \subsubsection{Adaptive group-sequential test for phase II clinical trials}

Sequential hypotheses tests are widely used in clinical trials applications \citep[see, for example,][]{jennison}. The most popular are so-called group-sequential methods, when samples of fixed size (groups) are drawn and analysed sequentially (interim analyses), allowing for early termination when  
sufficient information for acceptance or rejection of the hypotheses is collected.
There also exists a class of group-sequential methods called {\em adaptive}, when the size of the next group to be taken may depend on the results of interim analyses \citep{Dragalin}. In this way, adaptive group-sequential tests  are, in essence, the sequentially planned tests we consider in this paper.

In this subsection, we apply our technique  for  phase II clinical trials. We refer to the context of \cite{Fleming1982}, where a construction of group-sequential sampling plan is proposed for the clinical trials designed for testing  therapeutic effect of cancer treatments  based on the frequency of tumor ``regressions'' after the treatment has been applied. It is assumed that the frequencies are binomially distributed with the parameter $\theta$ representing the probability of regressions. The hypotheses of interest are $H_0: \theta\leq \theta_0$ vs. $H_1: \theta> \theta_1$, where $\theta_0<\theta_1$. As usual in applications, $[\theta_0,\theta_1]$ is considered an indifference zone, and we want to apply our optimal design in Subsection 4.1 to testing a simple hypothesis $H_0: \theta= \theta_0$ vs. $H_1: \theta= \theta_1$. We take as a reference the data of Table 12.1 in \cite{jennison}. 
\begin{table}[!t]
 \begin{tabular}{|c ||rc|rc|rc|rc||rc|}
 \hline
 &\multicolumn{8}{c||}{Max. 3 groups}&\multicolumn{2}{|c|}{Max. 5 groups}\\
 \hline
$\theta_0, \theta_1$ &\multicolumn{2}{|c|}{0.05, 0.2 } &\multicolumn{2}{c|}{0.1, 0.3}& \multicolumn{2}{c|}{0.2, 0.4}&\multicolumn{2}{c||}{ 0.3, 0.5}&\multicolumn{2}{|c|}{ 0.3, 0.5}\\
\hline
$\alpha$ &0.046& (0.046)&  0.050 &(0.063)&0.050 &(0.057) &0.050&(0.049)&0.051&(0.049)\\
$\beta$&0.09 &(0.087)& 0.10& (0.071)&0.10&(0.093)&0.10&(0.113)&0.10&(0.113)\\
ASN$_0$&34.1&(31.6)&23.6&(24.8)&30.8&(30.6)&36.3&(31.6)&36.0 &(31.6)\\
ASN$_1$&23.3&(26.8)&19.6&(21.6)&28.0&(30.4)&32.9&(35.5) &30.0&(35.5)\\
ANG$_0$&2.2&&1.8&&1.7&&1.8&&2.3&\\
ANG$_1$&1.8&&1.8&&1.8&&1.9&&2.7&\\
$\lambda_0$&154&&126.5&&199.8&&229.7&&230.2&\\
$\lambda_1$&57&&49.2&&69.8&&79.1&&69.1&\\
\hline
$n(\alpha,\beta)$&38.4&&31.8&&43.4&&49.9&&49.7&\\
\hline
$R_0$&1.13&&1.34&&1.41&&1.37&&1.38&\\
$R_1$&1.65&&1.62&&1.55&&1.52&& 1.66&\\
\hline
\end{tabular}
\caption{Adaptive vs. non-adaptive group-sequential tests for binary outcomes. The data in parentheses are taken from Table 12.1 in \cite {jennison} and correspond to the  plans in \citep[][ Table 2]{Fleming1982}} 
\end{table}
For each pair of  hypothesis points we applied the algorithm OTE of Subsection 4.1 taking as a cost function $c(m)=m,$ $m\in G=\{1,2,\dots, 40\} $, with $K=3$, taking into account that the plans of \cite{Fleming1982} use at most 3   groups. We used for our evaluations the value $\gamma=0.99$ for the weight coefficient, in order to  illustrate the effect of this input parameter  on the output performance of the plan. The choice of this parameter can be helpful with the ethical issue of clinical trials: if the treatment is turning out to be not efficient (which corresponds to rejecting $H_0$), it should be terminated as soon as possible (meaning the average sample number under $H_1$ should prevail when planning a real clinical trial). Large value of $\gamma$ is ment to make the average sample number ASN$_1$ under $H_1$ smaller in comparison with that under $H_0$, ASN$_0$. We use the grid size of $h=0.05$ for all the evaluations in this example.   

And we use the same nominal $\alpha=0.05$ and $\beta=0.1$ as in \cite{Fleming1982} to be able to compare the performance of the corresponding plans. The multipliers $\lambda_0$ and $\lambda_1$ are used, as intended, to comply with these requirements. 
A general-purpose gradient-free optimisation method by \cite{neldermeadarticle} was used to get as close as possible to the nominal values of $\alpha$ and $\beta$, with respect to the relative distance
\begin{equation}
 \max\{|\alpha(\psi,\phi)-\alpha|/\alpha,|\beta(\psi,\phi)-\beta|/\beta\}
\end{equation}
(the discrete nature of the binomial probabilities does not permit, generally speaking, to make the error probabilities exactly equal to $\alpha$ and $\beta$).

The fitted results are shown in Table 1. $ANG_0$ and $ANG_1$ are the values of the  average number of groups the optimal adaptive  plan takes under $H_0$ and $H_1$. The values of $\lambda_0$ and $\lambda_1$ are provided for reproducibility of the results.  $n(\alpha,\beta)$, exactly as above, is the minimum sample size a one-sample (FSS) test needs to comply with the error probabilities. Respectively, $R_i=n(\alpha,\beta)/ASC_i$ is the relative efficiency the optimal plan exhibits with respect to the non-sequential plan.
It is seen that the relative efficiency of the optimal sequentially planned test, with respect to the FSS test is higher than 1.5 in all the cases.

In general, we observe that our optimal adaptive (sequentially planned) tests, with as few as at most three groups allow to  save some 2 to 3 analyses (patients)  under $H_1$, on the average, in comparison with the plan of \cite{Fleming1982}.

To have an idea of the effect of using more groups, we ran the same fitting procedure using at most $K=5$ groups (last column in Table 1). It obviously is more efficient with respect to the one-sample plan (with the relative efficiency up to 1.66), and saves up to 5 analyses, on the average, in comparison with the 3-group plan.

Our program implementation in {\tt\small https://github.com/HOBuKOB-MEX/SPPRT} allows for virtually any number of groups (and any other parameters like  $\alpha$, $\beta$, etc.) when designing optimal plans for the binomial data.

\section{CONCLUSIONS}

In this paper, we proposed a method of construction of optimal sequentially planned tests. In particular, for  i.i.d. observations we obtained the form of optimal sequentially planned tests and formulas for computing their numerical characteristics like error probabilities, average sampling cost, average number of observations and the average number of groups.

A method of numerical evaluation of the performance characteristics is proposed
and computer algorithms of their implementation are developed.

For a particular case of sampling from a Bernoulli population, the proposed method is implemented in R programming language providing a computer code in the form of a public GitHub repository.

The proposed method is compared numerically with other known sampling plans.

A numerical comparison of the proposed tests with one-sample tests having the same error probabilities has been carried out. The relative efficiency  based on the average sampling cost compared to one-sample tests exhibits largely the same behaviour as that of the classical SPRT does, when the efficiency is based on comparison of  the average sample number with the FSS.

 \section*{ACKNOWLEDGEMENTS}

The author  gratefully acknowledges a partial support of SNI by CONACyT (Mexico) for this work.  

The author thanks the anonymous Reviewers and the Associate Editor for valuable comments and useful suggestions.

\end{document}